\documentclass[aps,prb,showpacs,groupedaddress,superscriptaddress,twocolumn]{revtex4}
\usepackage{amsmath,graphicx}

\begin{document}
\title{Magnetic Flux Control of Chiral Majorana Edge Modes in Topological Superconductor}
\author{Yan-Feng Zhou}
\affiliation{International Center for Quantum Materials, School of Physics, Peking University, Beijing 100871, China}
\affiliation{Collaborative Innovation Center of Quantum Matter, Beijing 100871, China}

\author{Zhe Hou}
\affiliation{International Center for Quantum Materials, School of Physics, Peking University, Beijing 100871, China}
\affiliation{Collaborative Innovation Center of Quantum Matter, Beijing 100871, China}

\author{Peng Lv}
\affiliation{International Center for Quantum Materials, School of Physics, Peking University, Beijing 100871, China}
\affiliation{Collaborative Innovation Center of Quantum Matter, Beijing 100871, China}

\author{X.C. Xie}
\affiliation{International Center for Quantum Materials, School of Physics, Peking University, Beijing 100871, China}
\affiliation{Collaborative Innovation Center of Quantum Matter, Beijing 100871, China}
\affiliation{CAS Center for Excellence in Topological Quantum Computation, University of Chinese Academy of Sciences, Beijing 100190, China}

\author{Qing-Feng Sun}
\email[]{sunqf@pku.edu.cn}
\affiliation{International Center for Quantum Materials, School of Physics, Peking University, Beijing 100871, China}
\affiliation{Collaborative Innovation Center of Quantum Matter, Beijing 100871, China}
\affiliation{CAS Center for Excellence in Topological Quantum Computation, University of Chinese Academy of Sciences, Beijing 100190, China}

\date{\today}
\begin{abstract}
We study the transport of chiral Majorana edge modes (CMEMs) in a hybrid quantum anomalous Hall
insulator-topological superconductor (QAHI-TSC) system in which the TSC region contains a Josephson
junction and a cavity.
The Josephson junction undergoes a topological transition
when the magnetic flux through the cavity
passes through half-integer multiples of magnetic flux quantum.
For the trivial phase, the CMEMs transmit along the QAHI-TSC interface as without magnetic flux.
However, for the nontrivial phase, a zero-energy Majorana state appears in the cavity,
leading that a CMEM can resonantly tunnel through the Majorana state to a different CMEM.
These findings may provide a feasible scheme to control the transport of CMEMs
by using the magnetic flux and the transport pattern can be customized by setting the size of the TSC.
\end{abstract}

\maketitle

\paragraph*{Introduction.}
Exotic excitations with characteristics of Majorana fermions in condensed-matter systems
are attracting widespread interest\cite{ReadN,QiXL1,Alicea1,Beenakker,ElliottSR}.
The chiral topological superconductors (TSCs) provide a fertile ground
in which the nature of such excitations can be explored.
Unlike the conventional superconductor, the chiral TSC can trap midgap Majorana zero modes,
bound to the superconducting vortices\cite{NKopnin,GVolovik} and electrostatic defects\cite{MWimmer}.
The Majorana zero modes obey non-Abelian statistics and can encode quantum information with potential applications
in fault-tolerant quantum computation\cite{ReadN,IvanovDA,CNayak}.
Moreover, the chiral TSC states can be classified by a Chern number $\mathcal{N}$
and have $\mathcal{N}$ chiral Majorana edge modes (CMEMs) residing at the edge\cite{QiXL1}.
The heterostructure formed by coupling the quantum anomalous Hall insulator (QAHI)
with an s-wave superconductor via the proximity effect can give rise to
the $p_x+ip_y$ pairing state providing a promising scheme to realize the chiral
TSC state\cite{QiXL2,WangJ}.
Experimentally, the QAHI state has been realized in Cr-doped\cite{ChangCZ1,Checkelsky,Kou,Bestwick,Kandala}
and V-doped\cite{ChangCZ2} $\mathrm{(Bi,Sb)_2Te_3}$ magnetic topological insulator thin films.
Very recently, He {\sl et al.} observed a half-integer conductance plateau at the coercive field
in a hybrid TSC-QAHI structure, which provides a hopeful signature of CMEMs in the chiral TSC\cite{HeQL},
and Zhang {\sl et al.} observed a quantized conductance plateau, which strongly supports the
existence of the Majorana state\cite{natureZhang}.

To further exploit the practical application of Majorana fermions in realistic devices,
it is a crucial step to effectively control and manipulate these Majorana modes.
Considering that Majorana fermion is a charge-neutral particle,
the direct effect on Majorana fermions by electric or magnetic methods should fail\cite{Park}.
As the Majorana zero modes are always located at the junction between topologically different domains
or tied to the defects, various kinds of schemes have been proposed to guide their positions
for braiding the Majorana zero modes\cite{Alicea2,MEzawa,BHeck,GFatin,PMarra,HWu}.
However, there remains a need for efficient methods to control and manipulate
the one-dimensional CMEMs\cite{YTanaka,YZhou}.

In this Letter, we propose a scheme to control the transport of CMEMs
in a hybrid QAHI-TSC-QAHI ribbon system, in which the TSC region contains
a line Josephson junction and a cavity pierced by a magnetic flux $\phi$, as shown in Fig.1(a).
The Josephson junction undergoes a topological phase transition when $\varphi$
($\varphi\equiv \pi\phi/\phi_0$ with $\phi_0=h/2e$) passes through half-integer multiples of $\pi$.
We calculate the Zak phase of the one-dimensional Josephson junction as a $Z_2$
invariant to distinguish the topologically different phases.
For $-\pi/2<\varphi<\pi/2$, the Josephson junction is trivial,
but for $\pi/2<\varphi<3\pi/2$, it is topologically nontrivial and a zero energy state
(Majorana state) exists in the cavity.
An incident Dirac electron from the left QAHI lead is
converted into a pair of Majorana fermions $\gamma_1$ and $\gamma_2$ at bottom left corner
of the TSC region [see Fig.1(a)].
For the trivial junction, $\gamma_1$ and $\gamma_2$ are transported to the left and right leads, respectively.
However, for the nontrivial junction, assisted by the tunneling between the Majorana state in the cavity
and CMEMs at the outer perimeter, both
$\gamma_1$ and $\gamma_2$ are transported to the left lead or right lead,
depending on the size of the TSC region.
This provides a feasible way to control the transport of CMEMs by using magnetic field.

\paragraph*{The model.}
In our setup, a TSC ring is interrupted by a Josephson junction (black dashed line),
enclosing a magnetic flux $\varphi$ and connected with two QAHI leads [see Fig.1(a)].\cite{note}
In the tight-binding representation, the low-energy physics of this QAHI-TSC-QAHI system can be described by the Hamiltonian
\begin{eqnarray}
 \mathcal{H} &=& \sum_\mathbf{i}\left[\psi_\mathbf{i}^{\dag} (T_0-\mu_{\mathbf{i}}) \psi_\mathbf{i}+(\psi_\mathbf{i}^{\dag}T_x\psi_{\mathbf{i}-\delta \mathbf{x}}+\psi_\mathbf{i}^{\dag} T_y \psi_{\mathbf{i}-\delta \mathbf{y}}) \right.\nonumber\\
 && \left.+ \Delta_{\mathbf{i}} c^{\dagger}_{\mathbf{i}\uparrow}
 c^{\dagger}_{\mathbf{i}\downarrow} + \mathrm{H.C.}
 \right],
\end{eqnarray}
with
$T_0=(m+4B)\sigma_z$, $T_x=-B\sigma_z-iA/2\sigma_{x}$ and
$T_y=-B\sigma_z-iA/2\sigma_{y}$. The regularization lattice constant $a=1$ and $\hbar=1$.
Here, $\sigma_{x,y,z}$ are Pauli matrices for spin,  $\psi_\mathbf{i}=(c_{\mathbf{i}\uparrow},c_{\mathbf{i}\downarrow})^T$, and
$c_{\mathbf{i}\sigma}$ and $c_{\mathbf{i}\sigma}^\dag$ are, respectively,
the annihilation and creation operators on site $\mathbf{i}$ with spin $\sigma=\uparrow,\downarrow$.
$\delta \mathbf{x}$ ($\delta \mathbf{y}$) is the unit vector along the $\mathbf{x}$ ($\mathbf{y}$) direction.
$\Delta_\mathbf{i}$ is the pairing potential and $\mu_{\mathbf{i}}$ is the chemical potential
which can be tuned by a gate voltage.
In the QAHI region, $\Delta_\mathbf{i} = 0$ and $\mu_{\mathbf{i}}=0$.
But in the TSC region, $\Delta_\mathbf{i} = \Delta$ due to the proximity effect by coupling of the
superconductor and $\mu_{\mathbf{i}}=\mu_s$.
$A$, $B$ and $m$ in Eq.(1) are material parameters, and the QAHI state with Chern number $\mathcal{C}=1$
can be realized with $m/B<0$.
Along the Josephson junction [the sites indicated by black dashed line in Fig.1(a)],
the pairing potential $\Delta_\mathbf{i}$ is set to zero and
the effect of the magnetic flux is included by Pierels substitution
$T_x\rightarrow T_xe^{i\pi\phi/\phi_0}$.\cite{RPeierls}
According to the phase diagram in Ref.[\onlinecite{QiXL2}], we set
$\left|\Delta\right|>\left|m\right|$ to make the TSC region in the chiral TSC phase
with $\mathcal{N}=1$.
It is worth noting that the chiral TSC was successfully realized experimentally\cite{HeQL}.
In the calculation, the parameters are set as $A=B=1.0$, $m=-0.2$, and $\Delta=0.8$.

\paragraph*{Controllable transport of CMEM.}
By using the nonequilibrium Green's function method, the normal tunneling, local Andreev reflection (LAR) and crossed Andreev reflecton (CAR) coefficients can be
obtained from\cite{SDatta,asun1,asun2,asun3}:
\begin{align} \label{TTT}
\begin{split}
   T(E) & =\mathrm{Tr}[\Gamma_{ee}^L G_{ee}^r\Gamma_{ee}^R G_{ee}^a], \\
   T^{\mathrm{LAR}}(E) & =\mathrm{Tr}[\Gamma_{ee}^L G_{eh}^r\Gamma_{hh}^L G_{he}^a],\\
   T^{\mathrm{CAR}}(E) & =\mathrm{Tr}[\Gamma_{ee}^L G_{eh}^r\Gamma_{hh}^R G_{he}^a],
\end{split}
\end{align}
where e and h represent electron and hole, respectively, $E$ is the incident energy.
$G^r(E)=[E-\mathcal{H}_{\mathrm{c}}-\Sigma^r_L-\Sigma^r_R]^{-1}$ is the retarded Green's function with
the Hamiltonian $\mathcal{H}_{\mathrm{c}}$ of the center region.
$\Gamma^{L/R}(E)=i[\Sigma_{L/R}^r-\Sigma_{L/R}^a]$ is the line-width function, with the self-energy
$\Sigma_{L/R}^r=\Sigma_{L/R}^{a\dagger}$ stemming from the coupling between the
left/right (L/R) leads and the center region\cite{DLee}.
Considering that there is only one edge mode in the QAHI leads,
the normal reflection coefficient can be obtained using $R=1-T-T^{LAR}-T^{CAR}$.

\begin{figure}
\includegraphics[scale=0.3]{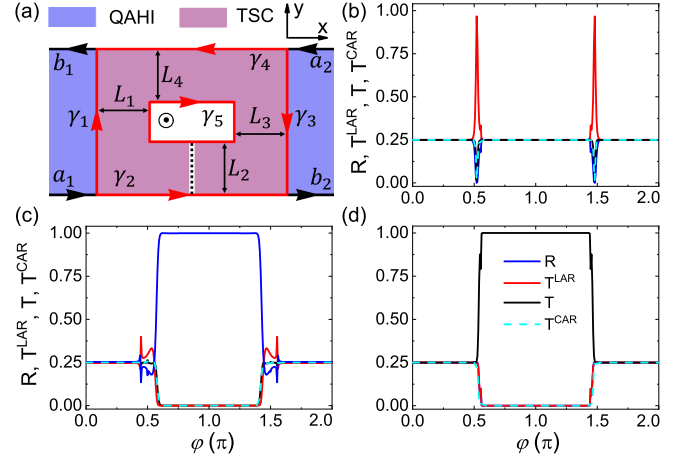}
\caption{
(a) Schematic diagram of the hybrid QAHI-TSC-QAHI ribbon system,
which can be obtained from a QAHI ribbon partially covered by an s-wave superconductor.\cite{HeQL}
The TSC region contains a line Josephson junction (black dashed line) and
a cavity pierced by a magnetic flux $\varphi$.
Black arrows label the QAHI edge states and red arrows indicate the CMEMs.
(b-d) The normal reflection coefficient $R$, LAR coefficient $T^{LAR}$,
normal tunneling coefficient $T$ and CAR coefficient $T^{CAR}$
as functions of $\varphi$ for three different size of the TSC region.
The chemical potential $\mu_s=0.8$, the size of the cavity $(L_x,L_y)= (201, 80)$, the distance (b)
${\vec{L}}\equiv (L_{1},L_{2},L_{3},L_{4})=(60, 60, 20, 20)$, (c)
${\vec{L}}\equiv(60, 20, 60, 20)$ , and (d) ${\vec{L}}\equiv(20, 60, 20, 60)$ in unit of $a$. The legends of (b) and (c)
are the same as in (d).
}
\end{figure}

Now, we turn to study the transport properties of the setup to show the scheme to control the CMEMs.
Figure 1(b-d) show the normal reflection coefficient $R$, LAR coefficient $T^{LAR}$,
normal tunneling coefficient $T$ and CAR coefficient $T^{CAR}$ in the zero-incident-energy case
as functions of the magnetic flux $\varphi$ for three different sizes of the TSC region.
In three cases, the size of the cavity is fixed,
and the distances $L_{i}$ ($i=1,2,3,4$) between the inner CMEM $\gamma_5$
and the outer CMEM $\gamma_i$ are changed.
When an electron propagating along the mode $a_1$ indicated by black arrow
from the left QAHI lead arrives at the trijunction,
it splits into two CMEMs $\gamma_1$ and $\gamma_2$ indicated by red arrows along
the outer boundary of the TSC region\cite{FuL1,YZhou,ZhangYT,AAkhmerov},
i.e., $a_1=1/\sqrt{2}(\gamma_1+i\gamma_2)$, as shown in Fig.1 (a).
Without the magnetic flux, eventually, $\gamma_1$ is backscattered to the left QAHI lead
as $\gamma_1=1/\sqrt{2}(b_1+b_1^{\dag})$, and $\gamma_2$ is transmitted to
the right QAHI lead as $\gamma_2=1/i\sqrt{2}(b_2-b_{2}^{\dag})$, respectively.
This implies that for the incoming mode $a_1$, the coefficients for normal reflection, LAR,
normal tunneling, and CAR are equal, so $R=T^{LAR}=T=T^{CAR}=1/4$ [see Fig.1(b-d)],
which is responsible for the observed two-terminal conductance $\sigma_{LR}=e^2/2h$
in He {\sl et al.}'s experiment\cite{WangJ,SChung,BLian,HeQL}.
When the magnetic flux is switched on, the CMEMs still transport in the above way
for $0<\varphi<\pi/2$ and $3\pi/2<\varphi <2\pi$ despite the size of the TSC region
as evident from Fig.1(b-d).
However, there is a significant change around $\varphi=\pi/2$ and $3\pi/2$.

\begin{figure}
\includegraphics[scale=0.33]{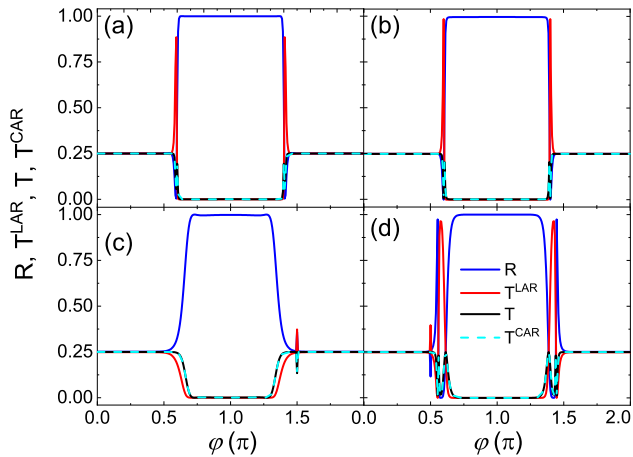}
\caption{ The transmission coefficients $T$, $T^{LAR}$, $T^{CAR}$ and $R$
as functions of $\varphi$ for (a) $\mu_s=0.8$, $\vec{L}=(60, 30, 60, 30)$;
(b) $\mu_s=0.8$, ${\vec{L}}=(60, 40, 60, 40)$;
(c) $\mu_s=0.2$, ${\vec{L}}=(60, 20, 60, 20)$;
and (d) $\mu_s=0.5$, ${\vec{L}}=(60, 20, 60, 20)$. The legends of (a-c)
are the same as in (d).}
\end{figure}

More interestingly, in the new phase with $\pi/2<\varphi<3\pi/2$,
the CMEMs behave differently, depending on the distances $L_{i}$.
Assuming that $L_{i}$ can be 20 and 60 representing short and long distances,
there are 16 ($2^4$) combinations of $L_{i}$ ($i=1,2,3,4$),
among which three transport patterns can be identified.
Here, we choose three representations to illustrate these patterns.
In Fig.1(b), $\vec{L}\equiv(L_{1},L_{2},L_{3},L_{4})=(60, 60, 20, 20)$,
the situation is same as the one without magnetic flux,
in which one CMEM ($\gamma_1$) is totally reflected,
and the other ($\gamma_2$) is transmitted entirely.
However, it can be observed in Fig.1(c) with $\vec{L}=(60, 20, 60, 20)$,
the perfect normal reflection dominates and $R=1$, with other processes disappearing absolutely,
including normal tunneling, local Andreev reflection, and crossed Andreev reflection.
In this case, both $\gamma_1$ and $\gamma_2$ return to the left lead and recombine as an electron.
This means that by tuning the magnetic field $\varphi$ through $\pi/2$ or $3\pi/2$,
the direction of the transport of CMEM $\gamma_2$ can be changed.
In the other case with $\vec{L}=(20, 60, 20, 60)$,
both $\gamma_1$ and $\gamma_2$ are transmitted to the right lead as an electron,
and the perfect normal tunneling
plays a leading role ($T=1$) with other processes prohibited as shown in Fig.1(d).
This means that the direction of the transport of CMEM $\gamma_1$ can be changed
when $\varphi$ passes through $\pi/2$ or $3\pi/2$.
At this point, it suggests
that the transport of the CMEMs can be controlled by tuning the magnetic flux.

Before analyzing the underlying mechanism of the scheme,
we pause to briefly discuss the availability of the scheme in different systemic parameters.
Taking the pattern in Fig.1(c) as an example, by changing the minimum of distances $L_{i}$,
the scheme is still valid as detailed in Fig.2(a-b).
Moreover, Fig.2(c) and (d) show that it also holds good for different chemical potential $\mu_s$
which can be changed by the gate voltage in real experiments.

\begin{figure}
\includegraphics[scale=0.3]{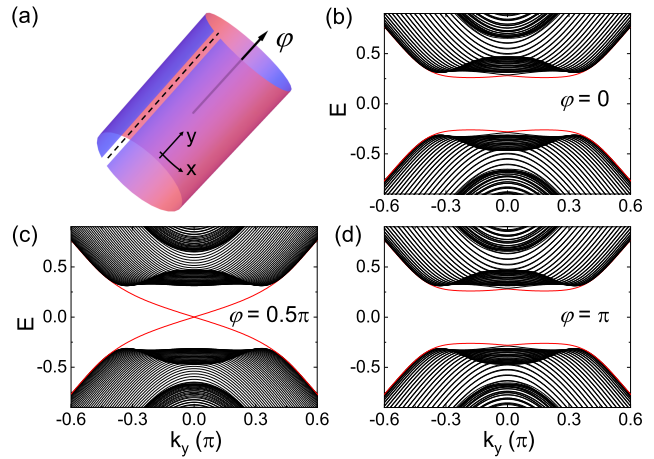}
\caption{
(a) Schematic diagram of a TSC Josephson junction in a cylinder geometry with a magnetic
flux $\varphi$ through its section cross.
(b-d) Band structure of the infinite cylindrical TSC Josephson junction with different
magnetic flux, (b) $\varphi=0$, (c) $\varphi=\pi/2$, and (d) $\varphi=\pi$.
The states denoted by red solid lines in the band
structure mainly reside near the junction denoted by black dashed line in (a).
The perimeter of cylinder $L_c=81a$ and all the other unmentioned parameters are the same
as in Fig.1.}
\end{figure}

\paragraph*{Topological transition.}
Above, we show that the coefficients $T$, $R$, $T^{LAR}$ and $T^{CAR}$ have a sudden change
when the magnetic flux $\varphi$ passes through half integer of $\pi$.
In order to explain this sudden change, we first study a TSC Josephson junction in a cylinder geometry with
a magnetic flux $\varphi$ through its section cross [see Fig.3(a)],
which is topologically equivalent to the TSC region containing a Josephson junction and a cavity [see Fig.1(a)].
The TSC in an infinite cylinder geometry is invariant under translation along the y axis,
so that the momentum $k_y$ is a good quantum number.
The band structures are shown in Fig.3(b-d) for different magnetic flux $\varphi$.
As can be seen in Fig.3(b), an energy gap exists for $\varphi=0$.
With the increase in $\varphi$ from zero, the gap gradually decreases with a
closing at $\varphi=\pi/2$ [see Fig.3(c)], and opens again when $\varphi>\pi/2$ [see Fig.3(d)].
Usually, the gap closing and reopening manifest a transition of the system between topologically different phases.\cite{MDiez,BWieder,FuL2}

\begin{figure}
\includegraphics[scale=0.31]{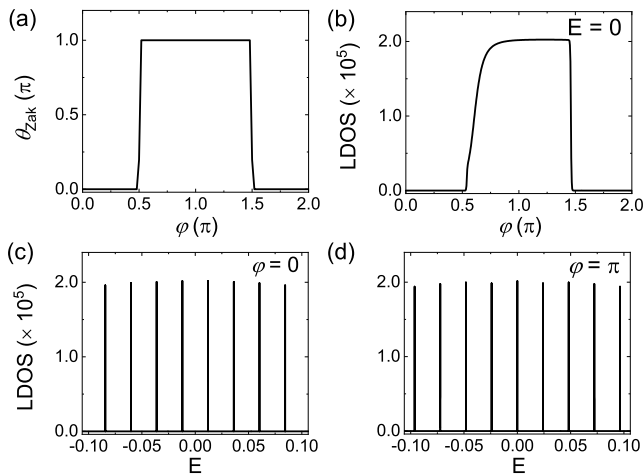}
\caption{ (a) Zak phase versus $\varphi$ in the infinite cylindrical TSC system [Fig.3(a)].
(b) $LDOS$ as a function of $\varphi$ for $E=0$ at the end of the finite cylindrical TSC system.
(c) and (d) The $LDOS$ at the end versus energy $E$ for $\varphi=0$ and $\varphi=\pi$, respectively.
Here $LDOS$ was summed over all the sites encircling the cylinder at the end,
i.e. $LDOS(E) = \sum_{j=1}^{L_c} LDOS_{0,j}(E)$ with the endpoint index 0 and the index $j$
along the perimeter. }
\end{figure}

To identify the distinct topological phases, we calculate the topological invariant
of the one-dimensional infinite cylindrical TSC Josephson junction.
For a one-dimensional system, the Berry phase across the whole Brillouin zone,
also called Zak phase,
\begin{equation}
  \theta_{Zak}=i\int^{\pi}_{-\pi}\langle u_{k_y}|\partial_{k_y}|u_{k_y}\rangle dk_y,
\end{equation}
can be used to characterize topological properties of the system\cite{JZak,MAtala,LFan},
where $u_{k_y}$ is the cell-periodic part of the Bloch function.
Moreover, the Zak phase can be related with the polarization difference,
which can be calculated from the sum over all occupied bands of the Wannier charge centers\cite{DXiao,SShen}.
Considering that the bands are highly degenerate [see Fig.3(b-d)],
we calculate the polarization difference by using the non-Abelian Berry connection\cite{RYu,DGresch}.
Figure 4(a) shows the Zak phase of the TSC Josephson junction as a function of magnetic flux $\varphi$.
The Zak phase can take only a value of zero or $\pi$ (modulo $2\pi$) for the system with inversion symmetry\cite{DXiao}.
For $\pi/2<\varphi<3\pi/2$, it is in the topological nontrivial phase with $\theta_{Zak}=\pi$,
and in the trivial phase with $\theta_{Zak}=0$ for the remaining parameter regime.

The physical characteristic of the nontrivial phase is the presence of zero energy Majorana state
at each end of a finite one-dimensional cylinder TSC system.
To pursue the end states, we evaluate the local density of states (LDOS) using the retarded Green's
function at a given site $\mathbf{i}$,
\begin{equation}
  LDOS_\mathbf{i}(E)=-\mathrm{Im}[\mathrm{Tr}G^r_{\mathbf{i},\mathbf{i}}(E+i\eta)]/\pi,
\end{equation}
where $\eta$ is a constant positive infinitesimal (in the numerical calculations, $\eta=10^{-6}$).
Figure 4(b) shows the LDOS at the end of the finite cylinder TSC system for zero energy.
The LDOS possesses a plateau for the nontrivial phase ($\pi/2<\varphi<3\pi/2$)
and vanishes for the trivial phase ($\varphi<\pi/2$ or $\varphi>3\pi/2$).
The LDOS decays with increasing the distance from the end and
is uniformly distributed around the cylinder (not shown here).
This confirms the existence of the zero-energy Majorana state at the end
in the nontrivial phase, which is consistent with the Zak phase.
Moreover, Fig.4(c) and (d) plot the LDOS at the end as functions of energy $E$
for $\varphi=0$ and $\varphi=\pi$.
Because the LDOS in both Fig. 4(c) and (d) shows a series of uniformly-spaced peaks, the discrete energy levels exist in both phases.
These discrete energy states stem from the quantum confinement on the CMEMs at the perimeter of the cylindrical TSC Josephson junction.
For $\varphi=0$, the energies of the states can be indexed as $E_q=qE_0$,
where $q$ takes on half-integer values and $E_0$ depends on the perimeter
of the cylinder $L_c$ ($E_0 \approx 0.024$ for $L_c=81$).
In this case, there is no zero-energy end state.
But $\varphi=\pi$, the $\pi$ Zak phase shifts $q$ to integer values
so that a zero-energy Majorana state emerges at the end of the cylinder TSC system.

Now, it is ready to revisit the results given in Fig.1(b-d).
Let us take the perfect normal reflection in Fig.1(c) as an example.
From above, we know that there is no zero-energy Majorana state in the cavity
when $\varphi<\pi/2$ or $\varphi>3\pi/2$.
In this trivial phase, CMEM $\gamma_1$ is backscattered to the left QAHI lead
and CMEM $\gamma_2$ transmits straightforwardly to the right QAHI lead [see Fig.1(a)],
resulting in that the four transmission coefficients are equal ($R=T^{LAR}=T=T^{CAR}=1/4$).
On the other hand, for the nontrivial phase with $\pi/2<\varphi<3\pi/2$,
a zero-energy Majorana state exists in the central cavity.
Now, CMEM $\gamma_2$ can resonantly tunnel across the zero-energy Majorana state in the cavity
into CMEM $\gamma_4$, then it goes back to the left lead, and combines
with $\gamma_1$ with $\gamma_1+i\gamma_4 =\sqrt{2} b_1$ as an electron
(i.e., $R=1$ and $T=T^{LAR}=T^{CAR}=0$), eventually.
Similarly, when $\pi/2<\varphi<3\pi/2$ in the parameter regime of Fig.1(d),
the resonant tunneling of CMEM $\gamma_1$ through the Majorana state in the cavity
into CMEM $\gamma_3$ occurs, leading to $\gamma_3+i\gamma_2 = \sqrt{2} b_2$, i.e.,
the perfect normal tunneling.
Therefore, based on the resonant tunneling of the CMEM,
the direction of the transport of the CMEM can be controlled by tuning the magnetic flux $\varphi$.

\paragraph*{Conclusion.}
In summary, we study the transport of CMEMs in the QAHI-TSC-QAHI system
with the TSC region containing a Josephson junction and a cavity.
With the change in the magnetic flux across the the cavity,
the Josephson junction undergoes a topological phase transition.
For the magnetic flux $\varphi<\pi/2$ or $\varphi>3\pi/2$,
the Josephson junction is in the trivial phase with the zero Zak phase,
and the CMEMs transmit along the QAHI-TSC interface as without magnetic flux.
However, for the magnetic flux $\pi/2<\varphi <3\pi/2$,
it is in the nontrivial phase with the Zak phase being $\pi$. In this case,
a zero-energy Majorana state exists in the cavity, leading to the occurrence of
one CMEM resonantly tunneling through the Majorana state to the other CMEM.
These findings may provide a feasible scheme to control the transport of CMEMs by using the
magnetic field and have potential applications for braiding the Majorana states.

\paragraph*{Acknowledgments.}
Y.F.Z. thanks Xiangru Kong and Yu-Hang Li for helpful discussions.
This work was financially supported by National Key R and D Program of China (2017YFA0303301),
NBRP of China (2015CB921102), NSF-China (Grants Nos. 11574007 and 11534001), and
the Key Research Program of the Chinese Academy of Sciences (Grant No. XDPB08-4).


\begin{thebibliography}{}
\bibitem{note}
The Josephson junction here is only for the purpose of continuous adjustment of
the magnetic flux in the cavity.
\end{thebibliography}
\end{document}